\documentclass[prl, preprint]{revtex4-1}
\usepackage{graphicx}
\usepackage{amsmath}
\usepackage{textcomp}

\usepackage[latin9]{inputenc}
\usepackage[francais,english]{babel}
\usepackage{graphicx}
\begin{document}

\title{MicroSQUID Force microscopy in a dilution refrigerator}
\thanks{P. C. and K. H. acknowledge support in the framework of
MICROKELVIN, the EU FRP7 low-temperature infrastructure under Grant No. 228464.
Z.S. W.
acknowledges thankfully the support of the French Embassy in Beijing and the CROUS.
This work has been supported by the French National
Research Agency, Grant No. ANR-09-Blanc-0211 SupraTetrafer and ANR- 09-BLAN-0146 SINUS.
W. Wernsdorfer and E. Eyraud are thanked for their advice and help during the cryogenic developments.
}



\author{D. J. Hykel, Z. S. Wang, P. Castellazzi, T. Crozes, G. Shaw, K. Hasselbach
}


\affiliation{
  Univ. Grenoble Alpes, Institut N\'eel, F-38042 Grenoble, France
  CNRS, Institut N\'eel, F-38042 Grenoble, France
    }
\email{klaus.hasselbach@neel.cnrs.fr}
\author{K. Schuster}
\affiliation{Institut de Radioastronomie Millim\'etrique,
 300 rue de la Piscine, Domaine Universitaire,
38406 Saint Martin d'H\`eres, France
}

\author{Z. S. Wang}
\affiliation{Beijing National Laboratory for Condensed Matter Physics,
  Institute of Physics, Chinese Academy of Science,
Beijing 100190, China 
}


\begin{abstract}
We present a new generation of a scanning MicroSQUID microscope operating in an inverted dilution refrigerator.
The MicroSQUIDs have a size of 1.21$ \ \mu$m\textsuperscript{2} and a magnetic flux sensitivity of
120 $\mu\Phi_{0} / \sqrt{\textrm{Hz}}$ and thus a
field sensitivity of 
550$ \ \mu \textrm{G}/ \sqrt{\textrm{Hz}}$.
The scan range at low temperatures is about 80 $\mu$m and a coarse displacement of 5 mm in x and y direction has been implemented. 
The MicroSQUID-to-sample distance is regulated using a tuning fork based force detection. A MicroSQUID-to-sample distance of 420 nm has been obtained. The reliable knowledge of this distance is necessary to obtain a trustworthy estimate of the absolute value of the superconducting penetration depth. 
An outlook will be given on the ongoing direction of development.
\keywords{SQUID \and Scanning Microscopy \and Superconductivty}

\end{abstract}
\pacs{ 85.25.Dq, 74.25.Ha} 

\maketitle

\section{Introduction}
\label{intro}

Scanning probe microscopy has been widely used for the study \cite{Kirtley2010} of superconductors, we briefly discuss advantages of each of the most common methods.
Magnetic force microscopy compared to other scanning techniques has the huge advantage of its high spatial resolution of $\sim$ 50 nm, and moreover recently significant progress has been made increasing its magnetic sensitivity \cite{auslaender2009}.
The vortex/tip interaction is not negligible and can lead to displacement of vortices.
By controlling the height of the MFM tip the displacement of vortices can be turned on and off and thus it can be used to control the vortex position. 
In order to interpret the MFM images quantitatively precise information of the tip magnetization is required.
Another scanning technique is the Hall probe microscopy \cite{bending2010},\cite{nishio2008}.
It combines good spatial resolution (850 nm) with a good magnetic resolution of
$\sim 2.9\times 10^{-3} {\textrm{G}} / {\sqrt{\textrm{Hz}}} $.
The Hall probe sensitivity is nearly independent of the Hall cross area (see \cite{kirtley99_review}), but is limited by Johnson noise, 1/f noise and the current density. Hall probe microscopes are very attractive for their operation in wide temperature and field ranges.


\section{Scanning SQUID microscopy}
As Superconducting Quantum Interference Devices (SQUIDs)  are flux sensors their field sensitivity varies linearly with the pick-up loop area. Therefore field sensitivity and spatial resolution can be optimized for the given experiment by choosing the size of the flux pick-up loop. 

For scanning SQUID microscopy (SSM) two different approaches are pursued for the design of SQUID sensors. 
The first approach consists in scaling down the size of highly sensitive conventional niobium tunnel junctions SQUIDs. The flux noise of these SQUIDs is of the order of 0.7 $\mu \Phi_{0}$/$\sqrt{Hz} $ which can be translated into a field noise of the order of 15 
$\mu {\textrm{G} /\sqrt{\textrm{Hz}  }  }$.
at 4.2 K. Operated without z-feedback loop, the SQUID chip is in contact with the sample surface \cite{koshnick2008}, making scanning at close distances difficult. The fabrication of multilayer Nb SQUIDs is very demanding and their wide Nb structures are prone to vortex pinning in magnetic fields of a few gauss, thus limiting the field-range of operation of these SQUIDs. If the SQUID is efficiently cooled, it is possible to image vortices in samples at temperatures of 100 K and more, as long as the sample is only in weak thermal contact with the SQUID \cite{kirtley_1999_APL}. 
The other approach consists in choosing a simpler SQUID design, aimed at increased spatial resolution. The most recent development is the SQUID on tip (SOT) design
 \cite{zeldov2010}, \cite{Single_spin}. Here a SQUID is deposited on the edge of a pulled quartz tube. SQUIDs with an effective diameter of less than 100 nm can be fabricated by three successive operations of evaporation. The quartz tube is scanned perpendicularly to the surface, in consequence the SQUID leads are parallel to the applied field, and the SOT can operate in fields up to 1 T \cite{Single_spin}. The SQUID readout relies on the use of a low temperature SQUID amplifier array conferring the detector a very high flux sensitivity of 50 n$\Phi_{0}$/$\sqrt{Hz} $ which can be approximated by a field sensitivity of $\sim$ 50 $\mu$G/$\sqrt{\textrm{Hz}}$ \cite{Single_spin} for a 160 nm diameter lead SOT. The SQUID-sample distance can be controlled by attaching \cite{1742-6596-400-5-052004} the metalized quartz tube to a tuning fork.

 \section{Scanning MicroSQUID microscope}
In the following we describe our approach of fabricating and using planar MicroSQUIDs, taking advantage of the power of lithography and achieving routinely high spatial resolution.
An array of hundreds of MicroSQUIDs is patterned by e-beam lithography on a silicon wafer. The Josephson Junctions are realized as Dayem-bridges of typically 20-30 nm width. The MicroSQUIDs are routinely made of aluminum by a lift-off process. After this step a photoresist mask defining the contour of each MicroSQUID chip is precisely aligned with the MicroSQUID array. This mask protects the MicroSQUIDs during the Si deep etch, which removes most of the Si in proximity of the MicroSQUID \cite{Hasselbach_IOP}. The precise etch simplifies the positioning of the MicroSQUID on the quartz tuning fork used for topographical imaging and height control \cite{Veauvy_RSI_2002}.
This geometry has the advantage of being mechanically robust, as the MicroSQUID-to-tip distance protects the MicroSQUID from the wear and tear of scanning,  allowing us to scan in the topographic feedback mode for several weeks. 

We use custom made room temperature electronics generating current ramps for the detection of the critical current of the MicroSQUID. As soon as the bias-current ramp reaches the critical current of the device, the bias-current is set back to zero in less than 25 nanoseconds. Once in the normal state an energy of the order of 4 $\times$ $10^{-15}$ joule is deposited in the MicroSQUID ($R_{N}$ =100 $\Omega$, $I_{c}$ of 40 $\mu$A). The deposited energy is easily removed by the connecting leads and the measurements can be repeated up to 10000 times per second. The present MicroSQUID has a flux noise of 120 $\mu\Phi_{0} / \sqrt{\textrm{Hz}}$ and can detect the stray field with a field noise of $ \sim$ 550 $\mu$G/${\sqrt{\textrm{Hz}}}$. In the following we will present the new implementation of our MicroSQUID microscope and show images validating the approach.

Our new microscope \cite{thesedanny},\cite{Zhao_thesis} is placed in an inverted dilution refrigerator \cite{sionludi} on top of a vibration isolated table. A single room temperature seal is sufficient to close the cryostat. The refrigerator is inverted in the sense that the sample space is on top of the refrigerator, approximately 40 cm above the table-top, a hundred liter helium dewar is suspended under the table supplying liquid He to the 4.2 K pot of the refrigerator. If the mixing chamber stage is unoccupied, cool downs to 4.2 K can be achieved in two hours and two hours more are sufficient to condense 35 liter of mixture, with a base temperature below 40 mK. Once the refrigerator is operating, the liquid helium consumption is less then 10 liters per day. 

\begin{figure}[h!]
 \centering
  \includegraphics[width=1\textwidth]{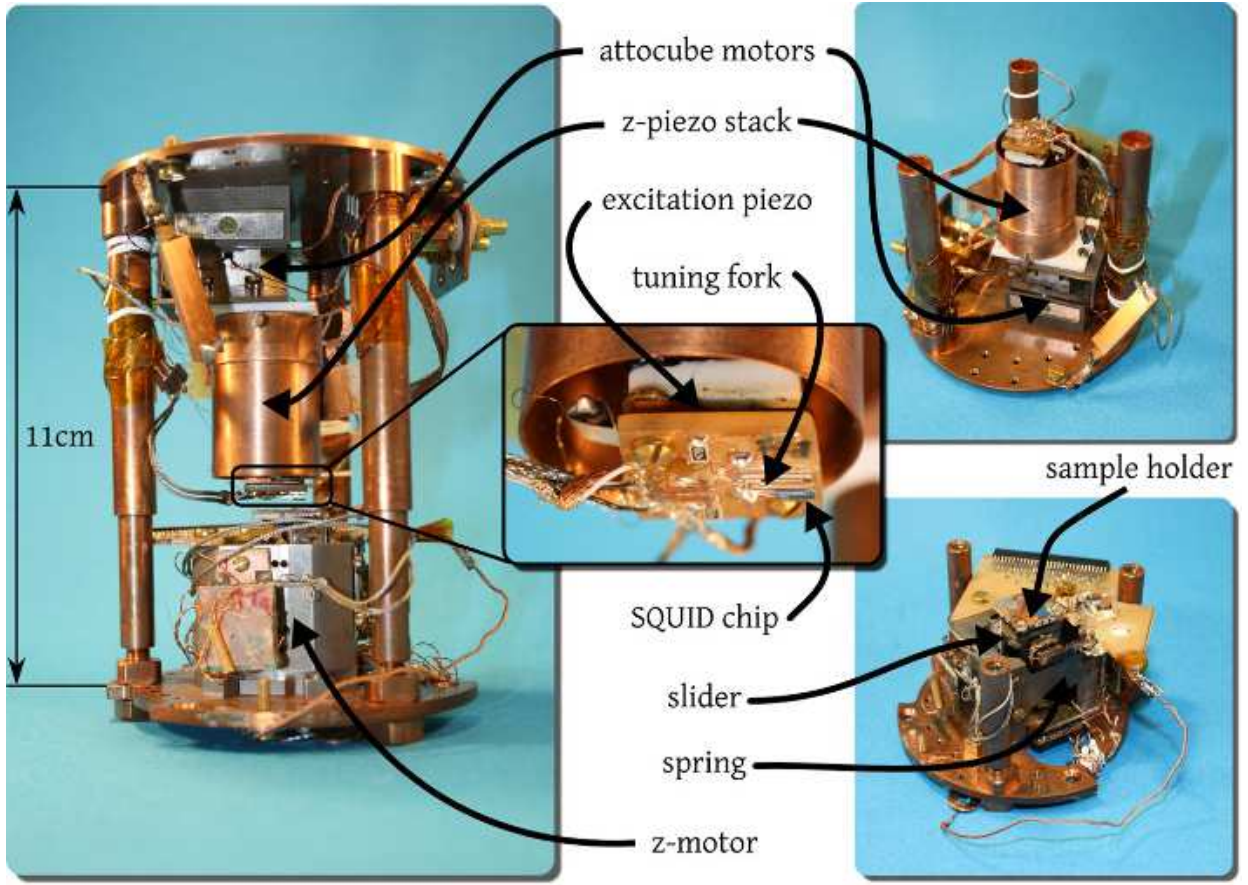}

 \caption{Photos of the microscope. Left: the completely assembled microscope, center: close-up of the tuning fork with the SQUID-chip underneath, right: the top part of the microscope with the sensors (top) and the lower part with the sample stage (bottom).}
 \label{fig:microscope_collage}
\end{figure}

The microscope has three piezo-electrically driven linear stages for coarse displacements along the xyz-directions of the order of 5 mm (Fig. \ref{fig:microscope_collage}). The z-direction stage encloses a hollow titanium prism containing a xy-scanner \cite{siegel95} allowing for a scan range of
80 $\mu$m in the x- and y-directions at 0.2 K, the base operating temperature. The piezo-electrical tuning fork is mechanically excited with a  "thickness" piezoceramic. At room temperature the tuning fork current is amplified with a custom-built current/voltage converter of gain $10^8$ V/A, and the voltage is demodulated with a fast lockin-amplifier \cite{lockin}. The phase signal of the lockin-amplifier enters a phase locked loop, changing the excitation frequency in order to maintain the tuning fork at resonance frequency. The difference of the actual resonance frequency from the set frequency enters a PID regulation loop regulating the distance between the SQUID tip and the sample surface. The distance is controlled by a z-piezo stack \cite{piezo-z} (C $\sim$ 1 $\mu$F) attached on the xy-stage \cite{attocube} and the regulation is implemented on a digital signal processing card \cite{sheldon}. The bandwidth of the regulation is about 1000 Hz, which must be degraded to 100 Hz at the lowest temperature in order to limit heating due to the charging of the z-piezo stack. The usual scan speed is 80 $\mu$m in 14 seconds. During a scan of 256 pixels 7680 critical current measurements are accumulated at 830 Hz which takes already 9 seconds out of the 14 seconds. The MicroSQUID is always thermally anchored to the mixing chamber, the sample stage on the scanner is only weakly thermally anchored to the mixing chamber or the still, allowing for fast thermal cycling of the sample above its superconducting transition temperature.

\begin{figure}[h!]
 \centering
 \includegraphics[width=0.35\textwidth]{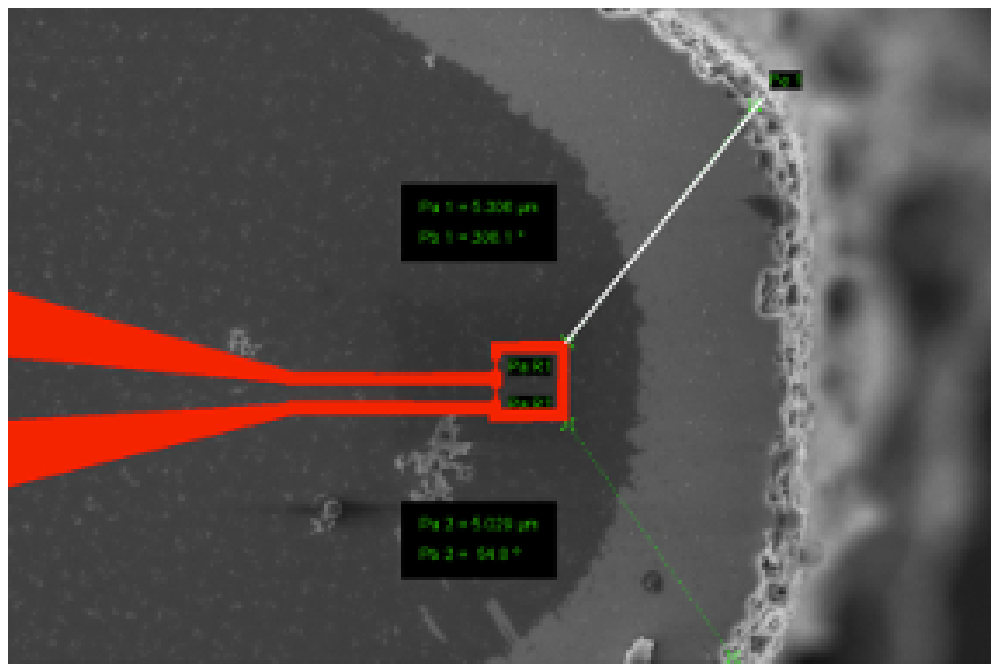}
 \includegraphics[width=0.25\textwidth]{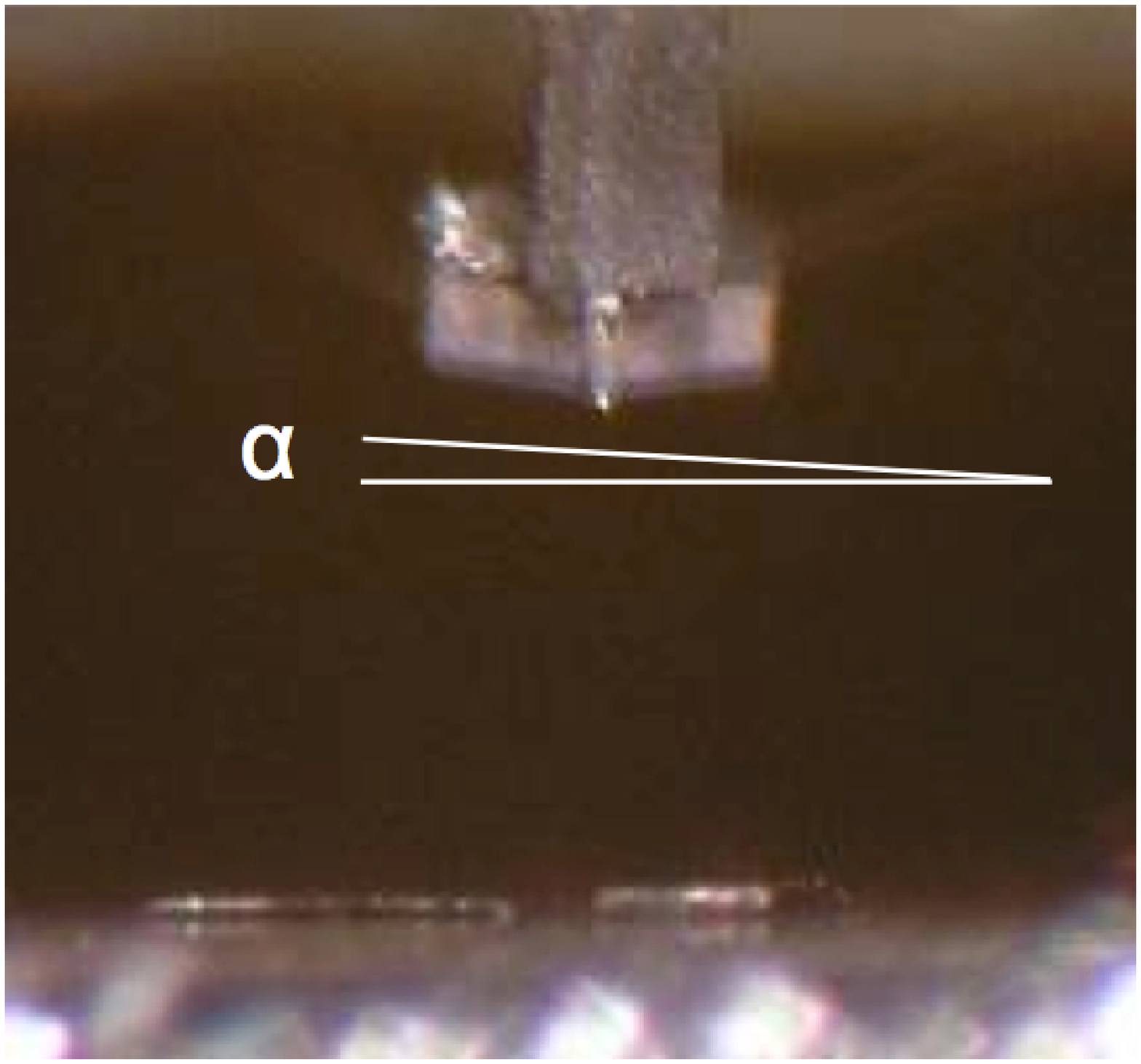}
 \includegraphics[width=0.35\textwidth]{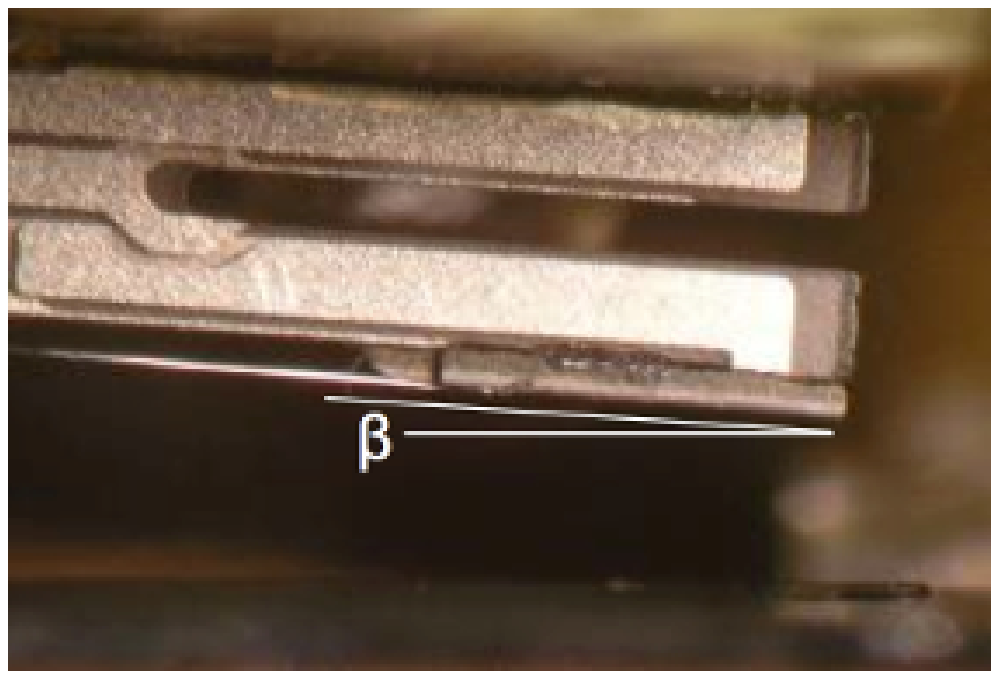}

\caption{Left:. SEM Image of a deep etched Si tip carrying a highlighted MicroSQUID having an inner diameter of $1\mu$m, the touch down point (white line) is selected in tilting the SQUID by 2$\textdegree$ sidewise (center) and 4$\textdegree$ downward (right), thus a MicroSQUID-to-sample distance of 420 $\pm$ 30 nm can be obtained.}
 \label{fig:alignment}
\end{figure}

Scanning magnetic probe microscopes can measure the magnetic stray field of a single vortex, and are thus employed to derive absolute values of the superconducting penetration depth. The absolute value of the penetration depth is of importance as it also enters in the expression of the penetration depth temperature dependence, which is used as a probe for the symmetry of the superconducting order parameter. 
The z component of the magnetic stray field of a vortex, $h_{z}$, measured in a plane at a height $z_{p}$, depends in the same manner \cite{Kogan} on this height and the penetration depth, $\lambda$:
\begin{equation}
 \frac{\partial h_{z} } {\partial \lambda} = - 2 \frac{\partial h_{z} } {\partial z_{p} }
\end{equation}
The reliable knowledge of the height $z_{p}$ is a prerequisite for deriving the superconducting penetration depth from the fitting of the field profile of a vortex.
In our case $z_{p}$ is the MicroSQUID-to-sample distance. The aim is to gain a precise control of this distance. A MicroSQUID-to-sample distance of 420 $\pm \ 30$ nm can be achieved under the condition that the MicroSQUID is placed at about 5 $\mu$m away from the touch down point (white line on the left panel of Fig. \ref{fig:alignment} and that the angles of approach are small and well controlled.

\begin{figure}[h!]
 \centering
\includegraphics[width=1\textwidth]{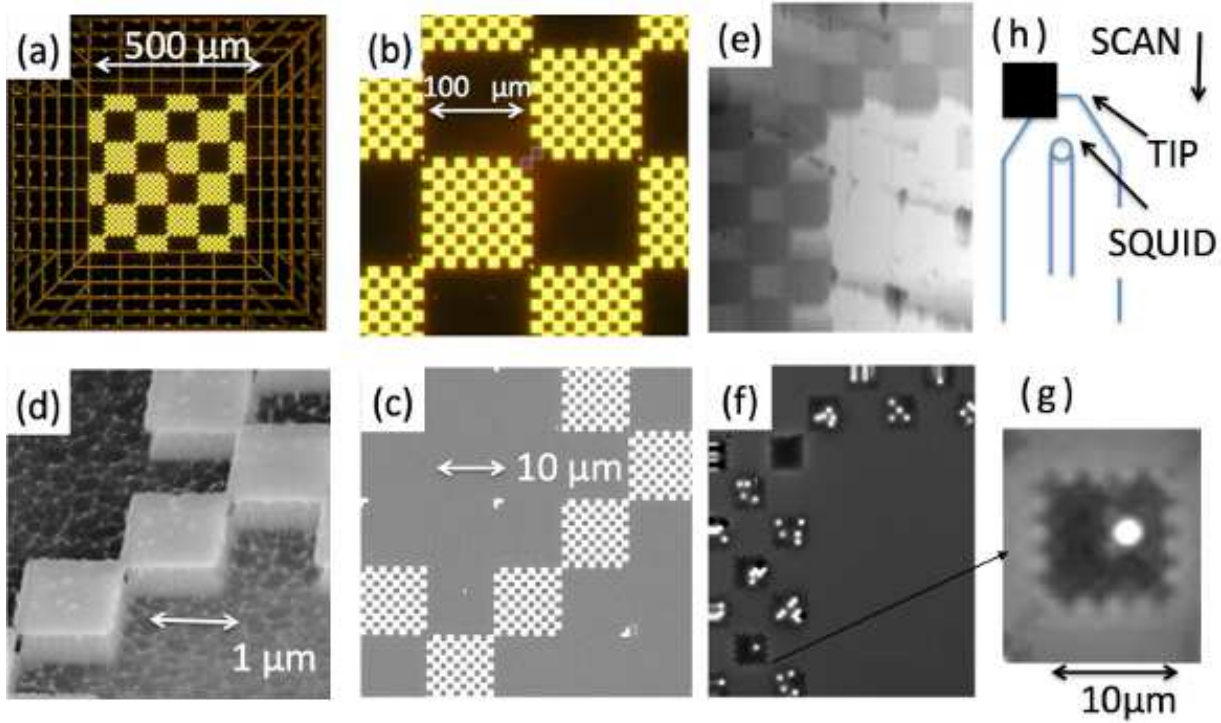}
 \caption{Sample: (a,b) optical images of the calibration pattern consisting of niobium squares of lateral length 100 $\mu$m, 10 $\mu$m and 1 $\mu$m. (c,d) SEM images: zoom on one small square of (b), (d) zoom on the smallest Nb square (1$ \mu$m), (e) is the topographic image obtained by recording the tuning fork signal, (f) is the simultaneously acquired magnetic image (T = 0.45 K) , (g) is the close-up view of a 10 $\mu$m square with one vortex in a 1 $\mu$m square, (h) schematized top view of the orientation of the probe-tip during scanning over the sample, represented by the black square, (not to scale).} 
 \label{fig:nb_sample}
\end{figure}

The other parameter entering the penetration depth determination is the displacement calibration of the xy-piezo-scanner. For this purpose we have designed a self-similar chessboard pattern of Nb squares of 100 $\mu$m, 10 $\mu$m, and 1 $\mu$m a side, alternating with empty squares of the same sizes. The pattern has been obtained by reactive ion etching of a 200 nm thick Nb film, giving also rise to topographic contrast. The pattern contains orientational marks (angles) indicating the direction to the center of the pattern (see Fig. \ref{fig:nb_sample}).

The topographic image Fig. \ref{fig:nb_sample}(e) of the calibration pattern is convolved with the shape of the probe-tip, (Fig. \ref{fig:alignment}, left), resulting in a softening of the edges of the 10 $\mu$m squares (dark grey) in the topographic image. As the scan angle is shallow, the tip does not touch the bottom of the pattern between the squares. The tip does touch the bottom of the pattern further away from the squares (clear grey). The magnetic image acquired simultaneously (Fig. \ref{fig:nb_sample}(f)) appears slightly shifted in upwards direction. The scan direction is from the top toward the bottom, and the tip is oriented towards the top of the image (Fig. \ref{fig:nb_sample}(h)), consequently the MicroSQUID senses first the magnetic contrast at a given point before the tip senses the topographic signal at this point, resulting in the observed shift in the images. The magnetic image is an image of the critical current of the MicroSQUID as function of position. In the magnetic image the ambient magnetic field gives rise to gray contrast, magnetic shielding at the location of the Nb squares appears dark and vortices, locations of high amplitudes of the stray field, appear bright. In looking closely at the close-up view of the 10 $\mu$m square, (Fig. \ref{fig:nb_sample}(g)), distinct 1 $\mu$m squares appear at the edge, surrounded by a bright halo, due to the compression of magnetic flux by the superconducting Nb.
From these calibration images we deduce an image size of 85 $\pm$ 1 $\mu$m for the vertical direction and 70 $\pm$ 1$\mu$m for the horizontal direction.


\section{Discussion and Conclusion}

MicroSQUIDs operation close to the quantum limit has been demonstrated \cite{koshnick2008}, \cite{Single_spin}, \cite{Hao_APL}. In order to obtain a measuring device the question of how to couple the magnetic signal of the sample into the SQUID becomes important. In static SQUID experiments it is possible to place magnetic nanoparticles as close as possible to nanometer wide constrictions in the superconducting loop, maximizing the flux coupling. In such a geometry the actual SQUID diameter is less relevant than the width of the constriction \cite{Cleuziou}, \cite{cond_mat}. 
The challenge of microscopy consists in scanning the probe above the surface and thus a minimal MicroSQUID-to-sample distance must be maintained. Moreover the smaller the pickup loop is, the closer the probe has to be in order to couple a significant amount of the flux source into the loop. The signal diminishes rapidly as function of MicroSQUID-to-sample distance for dipole, monopole and line sources of magnetic fields \cite{Kirtley2010}. A further difficulty arises from the fact that the microscope has to work in an unshielded environment at cryogenic temperatures. 

The microscope presented here has a tuning fork based height allowing us to scan several samples and calibration patterns during one cool-down lasting several month. Imaging at the base temperature of 0.2 K was achieved for the study of the ferromagnetic superconductor UCoGe \cite{Carley_UCoGe} . We showed that by controlling the shape of the very end of the MicroSQUID-tip we can reduce the MicroSQIUID-sample distance to at least 420 nm while scanning reliably, opening the path to the precise determination of the penetration depth in superconductors.

MicroSQUIDs have a huge potential for increase in sensitivity. We have recently reported on the continuous readout of hysteretic Al/Nb/W SQUIDs \cite{hazra:093109} in an magnetically unshielded environment, reaching a noise level of $5 \times 10^{-5} \Phi_{0} / \sqrt{Hz}$ or $\sim$ 130 $\mu$G/${\sqrt{\textrm{Hz}}}$ for a square MicroSQUID of 2.5 $\mu$m side length. This continuous readout will allow us to implement an on-chip feedback loop and to use highly sensitive SQUID electronics for the MicroSQUID-readout.
We are currently developing a low temperature sample displacement (T $<$ 0.1 K) mechanism, driven by room temperature motors, adapted for a novel inverted dilution refrigerator, built and tested recently. This design will remove most of the piezoelectric high voltage lines from the SQUID environment, reducing possible sources of noise and leveling the path to user friendly low temperature scanning MicroSQUID microscopy.

\bibliographystyle{spmpsci}  

\bibliography{Jltp_3.bib}
%
%

\end{document}